\documentclass[preprint]{ptptex}

\usepackage{wrapft}

\def\beqra{\begin{eqnarray}}
\def\eeqra{\end{eqnarray}}
\def\beqast{\begin{eqnarray*}}
\def\eeqast{\end{eqnarray*}}
\def\be{\begin{enumerate}}
\def\ee{\end{enumerate}}

\def\beq{\begin{equation}}
\def\eeq{\end{equation}}

\notypesetlogo  %comment in if to eliminate PTPTeX logo
\title{%        %You can use \\ for explicit line-break
Possible Chandra-Mechanism for Generation Bound\\
}
\author{%       %Use \sc for the family name
Kazuo {\sc Koike}% {\sc Tomonaga}\footnote{A friend of Schwinger 
}
\inst{%         %Affiliation, neglected when [addenda] or [errata]
Department of Natural Science,~Kagawa University,
Takamatsu 7608522, Japan
\\
}
\recdate{July **, 2004%      %Editorial Office will fill in this.
}
\abst{
In the generation structure, the quark mass increases extremely rapidly with the increase of generation index, and there is the bound for generation number. The ground for this bound is investigated on the basis of a certain kind of composite model of leptons and quarks, in which they are supposed to be composed of sub-constituents with fermi statistics. Possible Chandrasekhar-like mechanism for generation bound is proposed.
}
\begin{document}

\maketitle

\noindent
{\it 1. Introduction }

The masses of leptons and quarks extremely rapidly increase accompanied with the generation number. Especially, it is surprising that the top-quark mass is larger than the u-quark mass about a factor $\rm 10^4$.  That is, the u quark mass is estimated as a few MeV order by making use of current algebra,\cite{rf:quark_mass}  while the mass of top quark is about 178 GeV. The other remarkable fact is the existence of bound for number of generations. Is it possible to understand these facts in a unified way? This paper is concerning to this problem.

It is known that the number of generation is restricted\cite{rf:GN3}  to just 3 so far as the neutrino masses are bound below $\rm m_Z/2$. This 3 generation structure is explained by choosing an appropriate symmetry group as the initial condition in the standard gauge theory or GUTs. The symmetry concerning to the repeated generation structure called as the ``horizontal symmetry" is badly broken, because the quark mass increases extremely rapidly with the increase of generation number.
Several  theoretical  approaches   have   suggested   the 
existence of upper-bound of generation  number.  Asymptotic  free 
condition\cite{rf:Gross} in QCD restricted it below 8. Nishijima  has
shown\cite{rf:Nishijima} 
that color  confinement  condition  based  on  BRS  transformation 
restricts it below 4. On the  other  hand,  generation  structure  
has been investigated on the basis of the exceptional  group,  by 
making use of finiteness of rank of this group. Especially,  it 
should be noted that the quasi Nambu-Goldstone fermion model\cite{rf:Yanagida} of 
leptons and quarks based on $ E_7/SU(5) \times SU(3) \times U(1)$ gives at most the 3 generation structure.\cite{rf:Kugo-Y}
Though these approaches are very interesting and important, it should be noted that these are extensive approach based on the consistency condition for realizing the object. 
As a complementary approach, we will investigate in this paper the problem of generation bound on the basis of a certain kind of composite model of lepton and quarks. The basic motivation leading to our approach is that

\noindent
~~(1) the quark mass increases extremely rapidly with the increase of generation
number,

\noindent
~~(2) there is the bound for generation number.

 \noindent
 As a proto type, we will construct a model concerning to this problem on analogy of Chandrasekhar mechanism.\cite{rf:Chandra}

\noindent
{\it  2. Composite model and sub-constituent with fermi statistics}
 
It is  known that Chandrasekhar have ever shown that the white dwarf has its upper bound of mass\cite{rf:Chandra} due to the pressure of degeneracy of electron. We will take a composite model of leptons and quarks in which they are composed of a certain kind of sub-constituents obeying the fermi statistics. As the constructive force of leptons and quarks, the very short-range effective force acting on only the mass of sub-constituents with specific quantum number is assumed.  That is, our system is as is the gas of sub-constituents obeying fermi-statistics, and it is support by degenerate pressure. The generation number will be assumed, for the time being, to be given by generation suffix of sub-constituent and the mass will increase extremely rapid according to the increase of generation number. The derivation of upper bound of mass of composite system is similar to the Chandrasekhar's one.

\vspace{.25cm}
\noindent
{\it 3. Equilibrium equation and mass of composite system }

In first, we will introduce the very short-range effective force acting on only the mass of sub-constituents. In the center of mass system of composite system, we will suppose the force given by Yukawa potential for unit mass,

\begin{equation}
V(r) = -~ G_s \frac{e^{-\mu r}}{r} ,
\label{eq:Potential}
\end{equation}

\noindent
where $G_s$ and $\mu$ are supposed to be extremely large to realize short range strong force.
For small $r$, Eq.~(\ref{eq:Potential}) is represented as,

\begin{equation}
V(r) = - \frac{G_s}{r} ,
\label{eq:Short}
\end{equation}

\noindent
It should be noted that  Eq.~(\ref{eq:Short}) is the same form as ordinary potential of gravity except $G_s$.
According to Chandrasekhar\cite{rf:Chandra}, we will assume spherical symmetric composite system which is characterized by pressure $P(r)$ and density distribution $\rho(r)$.
The dynamical equilibrium condition between inner and outer forces gives the following equation,

\begin{equation}
\frac {d P(r)}{d r} = - {\frac { G_s M(r)}{r^2}} \rho(r) ,
\label{eq:equilibrium}
\end{equation}

\noindent
where $M(r)$ represent the quantity of matter inside the surface with radius $r$, 

\begin{equation}
M(r) = \int_0 ^r 4\pi \rho(r)r^2 dr.
\label{eq:total_mass}
\end{equation}

In order to resolve equilibrium equation Eq.~(\ref{eq:equilibrium}),the other equation of state is required. In the case of star, the polytrope equilibrium based on Emden's model is often assumed. In our case, there is no reason on this equilibrium. Then, we will simply replace Eq.~(\ref{eq:equilibrium}) and Eq.~(\ref{eq:total_mass}) as, 
\begin{equation}
\frac {P_c}{R} \sim  \frac {G_s M \rho_c}{R^2},~M \sim \frac {4 \pi}{3} R^3\rho_c ,
\label{eq:equilibrium-approxi}
\end{equation}

\noindent
where $P_c$ and $\rho_c$ represent the pressure and density in the center of composite system. It should be noted that this treatment corresponds to the case when the gradient of pressure is uniform in composite system.
From Eq.~(\ref{eq:equilibrium-approxi}), we have
 
\begin{equation}
M \sim \left[~\frac {1}{G_s^3}~ \frac {P_c^3}{\rho_c^4}~\right]^{1/2}.
\label{eq:mass-result}
\end{equation}

\noindent
In the case when the polytrope equilibrium is assumed, it is known that

\begin{equation}
M_n = \left[f_n ~\frac {1}{G_s^3}~ \frac {P_c^3}{\rho_c^4}\right]^{1/2}
\label{eq:mass-polytrope}
\end{equation}

\noindent
instead of Eq.~(\ref{eq:mass-result}), where $f_n$ is the polytropic index of magnitude of order 1.

%\break
\vspace{1cm}

\noindent
{\it 4.The pressure of degenerate fermi gas}

In the cubic space with the side length $L$, the number of fermion $N$ below fermi surfaces given as

\begin{equation}
N = \frac {8 \pi \int_0^{P_F} p^2 dp} {(h/L)^3}.
\label{eq:total_number}
\end{equation}

\noindent
Then the number density of fermion is given as
\begin{equation}
%n = \frac{N}{V} = \frac {1}{\pi^2 \hbar^3} \int_0^{P_F} p^2 dp.
n = \frac{N}{V} = (\pi^2 \hbar^3)^{-1} \int_0^{P_F} p^2 dp.
\label{eq:number_density}
\end{equation}

\noindent
From Eq.~(\ref{eq:number_density}), the fermi momentum $P_F$ is given as

\begin{equation}
P_F =  (3\pi^2 \hbar^3 n)^{1/3}.
\label{eq:P_F}
\end{equation}

In the relativistic case in which the sub-constituents fermions have sufficiently high energy, the energy of sub-constituents $\epsilon$ is $\epsilon = cp$ for sufficiently large momentum, then the collective energy of $N$-body system is given as

\begin{equation}
E = V~ (\pi^2 \hbar^3)^{-1}  \int_0^{P_F}(cp)~ p^2 dp, 
\label{eq:collective_energy}
\end{equation}

\noindent
then, the degenerate pressure is given as

\begin{equation}
P = - \frac {dE}{dV} = (3\pi^2)^{1/2}\left(\frac{\hbar c}{4}\right) n^{4/3}.
\label{eq:pressure}
\end{equation}

The density $\rho$ is related to the number density $n$ by

\begin{equation}
\rho = m_s n ,
\label{eq:density}
\end{equation}

\noindent
where $m_s$ is effective mass associated with the sub-constituents. In the case of degenerate star, it is given by the nucleon mass associated with one electron.

\vspace{.25cm}
\noindent
{\it 5.  Possible Chandra mechanism in composite model }

From Eq.~(\ref{eq:pressure}) and Eq.~(\ref{eq:density}), the degenerate pressure $P$ is is given as

\begin{equation}
P = (3\pi^2)^{1/2}\left(\frac{\hbar c}{4}\right) m_s^{-4/3} \rho^{4/3}.
\label{eq:pressure2}
\end{equation}

\noindent
Then, from Eq.~(\ref{eq:mass-result}) it is seen that the upper mass supported by degenerate pressure does not depend on density $\rho_c$. The explicit representation of this upper mass of our composite system is 

\begin{equation}
M_{s} \sim \left[~ \frac {(3\pi^2)^{3/2}(\hbar c/4)^3 }{G_s^3 m_s^4}~~\right]^{1/2}.\label{eq:upper_mass}
\end{equation}

\noindent
{\it 6. Model of the generation bound}

In order to see the possible applicability to composite model of leptons and quarks, we will estimate the magnitude of our coupling constant $G_s$.
It is surprising that the mass of particles of the third generation is extremely large in the view of symmetry breaking. Especially, it should be noted that the mass of top quark $ t$ is extremely larger than  charm quark $ c$ about factor $10^2$. 
For simplicity, we will suppose the hidden mass of fourth generation up-quark is large than $ t$ about factor  $10^2$ interpolating the previous factor. Thus, we will take, for example, $\rm M_{s} \sim 10~ TeV$ and $\rm  m_s \sim m_{top}$,
 then 

\begin{equation}
G_{s} \sim 1.17 \times 10^{-5}~ \hbar c~ (GeV/c^2)^{-2}.
\label{eq:estimate_mass}
\end{equation}

\noindent
This value seems to be not so extreme one, compared with some other coupling constants.\footnote{
\noindent
Fermi coupling constants $G_F$ is given as\cite{rf:quark_mass}
\begin{equation*}
G_{F} = 1.16  \times 10^{-5}~ (\hbar c)^3~ GeV^{-2}.
\label{eq:G_F}
\end{equation*}
}
\noindent
However, this coupling constant $\rm G_s$ is very large compared with the gravitational constant\cite{rf:quark_mass}
 $G_N$, which is known to be extremely small,\footnote{The validity of gravitational inverse-square law is confirmed only for the distance larger than \indent~~~ $1 mm$ in the present stage.\cite{rf:Newton}}
\begin{equation}
G_{N} = 6.71  \times 10^{-39}~ \hbar c~ (GeV/c^2)^{-2}.
\label{eq:G_N}
\end{equation}

Thus, the existence of upper mass bound is shown in our composite system.
Under the existence of such upper bound of mass of composite system, what phenomena will appear when large energy above the upper bound is localized. In the case of star in AGB phase, it is known to emit its mass, and gradually move to white dwarf.
In our composite system, a certain kind specific behavior will be observed when the energy above such upper mass limit is localized. This specific behavior or a certain enhancement may be observed in  processes such as the multiple production of $t\bar t$ pairs. 
Further, a certain kind of rare processes accompanied with the crash of the composite system may also be observed.

%Possible indications in our model
%degenerate star
%It is surprising that the masses of particles of the third generation is extremely large in the view of symmetry breaking. Especially, it should be noted that the mass ratio of top quark $ t$ to up quark $ u$ is about $10^5 : 1$.

\vspace{.25cm}
\noindent
{\it 7. CKM matrix and quark mass}

The other difference will appear in the relation between the CKM matrix and the quark masses.
In the standard model based on the gauge theory of elementary (not composite) particles, CKM matrix is in principle determined by diagonalization of quark mass matrices. In semi-empirical approach to this problem, derivation of the extreme large mass of top quark is often difficult.\cite{rf:Koike-mass} Moreover, the important problem why the u-quark mass is turned from the other leptons and quarks and slightly smaller than the mass of d-quark\cite{rf:quark_mass} (the Cleopatra problem in leptons and quarks) is essentially hidden in merely  the level of tuning of big-quark mass, because the light-quark mass is about $10^{-4}$ order of the big-quark mass.

In a certain kind of composite model of leptons and quarks, $\rm SU(2)_L \times U(1)$ electroweak structure may be introduced in sub-system under a certain kind of conditions. In such case, the quark mass in our model will reflect both the diagonalization of mass matrix of sub-system and the mass concerning to our compositeness.
Details of this problem will be discussed elsewhere.
 
\vspace{.25cm}
\noindent
{\it 8. Discussion}

  In this paper, we have investigated a possibility to interpret the existence of bound for generation number supposing a mechanism similar to the Chandrasekhar's one for degenerate star. We have taken a composite model of leptons and quarks which sub-constituent obeys fermi statistics, and assumed short range Yukawa-type effective potential in Eq.~(\ref{eq:Potential}). Our model have shown formally the existence of upper bound of mass of leptons and quarks provided that the coupling constant $G_{s}$ is  appropriately chosen.  The magnitude of estimated coupling constant $G_{s}$ is very large compared with the one of the gravitational constant $G_N$, however, there is no problem so far as it represents the strength of the effective interaction. 
Can this force be related to the fundamental force?
%It is terrible to introduce such force. 
In such case, in order to preserve  short range property of this force, the $\mu$ parameter in Eq.~(\ref{eq:Potential}) should be sufficiently large. 
In the short range correction, C meson (Concentrate meson) theory\cite{rf:Sakata}  introduced to cancel the ultra-violet divergence should be remembered.  Our short range force may be concerned with such approach in anywhere.
For further investigation of peculiar short range force, a certain force appearing in non-linear wave equation\cite{rf:NL}, for example, is interesting. 

We have supposed composite system composed of fermion sub-constituents. One of remarkable models of leptons and quarks is the rishon model.\cite{rf:rishon,rf:KK} It is known that almost all quantum number except the electric charge loses its meaning in the extreme space time condition such as black hole. It should be noted that the rishon model is  based on the simple concept of the elementarily of electric charge, and in this sense it is natural. In our composite model, the fundamental particles are composed of sub-constituents with fermi statistics.  Can the rishon be the sub-constituents of our model? In the nuclear physics, it is known that various mode such as shell model mode or collective mode appears together in a few body system. Then, the rishon model itself can be the model of our composite system. Alternatively, if we modify this model to Thomson-type model,\cite{rf:Tanaka} where many constituents is distributed in a charge compensating the charges of object, our picture will hold with naturally. Possible another modification is such that a rishon is an assembly of many fermions and the electro charge is carried by a representative fermion.  This type model will be formally introduced, for example, in such a way that the color "degenerates", that is, not distinguished in exclusion principle, in a model similar to Pati-Salam's one.\cite{rf:Pati-Salam} 

In this paper,we have shown that existence of super-heavy composite particle are impossible by Chandrasekhar-like mechanism provided that short range effective force is introduced. In our approach, possible effect of quantum fluctuation is already taking into account in the derivation of fermi momentum $ P_F$ in Eq.~(\ref{eq:P_F}). Investigation of problems concerning to further short-range effect, the validity of large coupling constant $\rm G_s$ and the structure of our composie system together with covariant formulation is further problem.
Finally, in the development of our model where our effective force is related the fundamental force with universal interaction, it is interesting to investigate possible existence of rare process to emit some energy flowing into  braneworld\cite{rf:Akama} etc.

%\section*{Acknowledgments}
\vspace{0.25cm}
\noindent
{\it Acknowledgments}:~
This work was supported by Grant-in-Aid of Japanese Ministry of  Education, 
Science, and Culture (15540384).

%\section{Remarks}
%notation
%\parskip=5mm %$m_i^{(s)}$ %\par %$m_i^{(b)}$
%\subsection{Fractions}
%\section{References}
%\section*{Acknowledgments}
%We would like to thank ...........
%\appendix
%\section{First Appendix} %Empty argument \section{} yields `Appendix'. 
%\section{Second Appendix}

\end{document}